\documentclass[conference]{IEEEtran}
\IEEEoverridecommandlockouts    

\usepackage{multirow}
\usepackage{lipsum}
\usepackage{amsmath}
\usepackage{amssymb}
\usepackage[ruled,vlined]{algorithm2e}
\usepackage{graphicx}
\graphicspath{{./Figures/}}
\usepackage{url}

\begin{document}

\title{\LARGE \bf Dynamic Risk Generation for Autonomous Driving: \\
Naturalistic Reconstruction of Vehicle-E-Scooter Interactions}

\author{
Abin Mathew, Zhitong He, Lingxi Li, and Yaobin Chen\thanks{* Corresponding author: Yaobin Chen.}\\
\emph{Elmore Family School of Electrical and Computer Engineering, Purdue University, Indianapolis, USA}\\
\{mathe102, he733, lingxili, chen62\}@purdue.edu
}

\maketitle
\thispagestyle{empty}
\pagestyle{empty}


\begin{abstract}
    The increasing, high-risk interactions between vehicles and vulnerable micromobility users, such as e-scooter riders, challenge vehicular safety functions and Automated Driving (AD) techniques, often resulting in severe consequences due to the dynamic uncertainty of e-scooter motion. Despite advances in data-driven AD methods, traffic data addressing the e-scooter interaction problem, particularly for safety-critical moments, remains underdeveloped. This paper proposes a pipeline that utilizes collected on-road traffic data and creates configurable synthetic interactions for validating vehicle motion planning algorithms. A Social Force Model (SFM) is applied to offer more dynamic and potentially risky movements for the e-scooter, thereby testing the functionality and reliability of the vehicle collision avoidance systems. A case study based on a real-world interaction scenario was conducted to verify the practicality and effectiveness of the established simulator. Simulation experiments successfully demonstrate the capability of extending the target scenario to more critical interactions that may result in a potential collision.
\end{abstract}
\section{Introduction}
\label{sec:introduction}
The expansion of micromobility has transformed urban mobility with an increasingly popular option for “last-mile” travel. Although mobility brings convenience and relief from traffic congestion, especially in heavy traffic urban areas, the crash reports \cite{cpsc-crash,ucsf-crash} demonstrate a significant rise in injuries, particularly among riders on e-scooters, highlighting the non-negligible risk for daily users. The transformation of novel artificial intelligence technique, especially the data-driven vehicle control methodology and the application of foundation models \cite{cui2024llm}, leverages the traffic data to pursue a higher level of automation and intelligence for ground vehicles. Researchers have emphasized the necessity to include vulnerable road user (VRU) safety concerns under the development of the new transitions \cite{Libook,TIV}. 

A recent study \cite{seitakari2025comparing} demonstrates the differences between e-scooter and bicycle rider injuries, with more head injuries observed in e-scooter riders. Around \(40\%\) of the e-scooter injuries happened at night due to limited vision. Traditional VRU testing scenarios occasionally include e-scooters. Furthermore, among a limited number of recent traffic datasets that include e-scooters in urban traffic scenarios,  a majority of them focus more on object detection \cite{yan2025machine, chen2024performance} and intention prediction \cite{zhang2024intent} rather than investigating the vehicle's reaction to dynamic e-scooter interaction behaviors. Therefore, a unified testing benchmark is urgently needed to evaluate the interaction between the vehicle and the e-scooter rider. 

\begin{figure}[t]
\centerline{\includegraphics[width=1.0\linewidth]{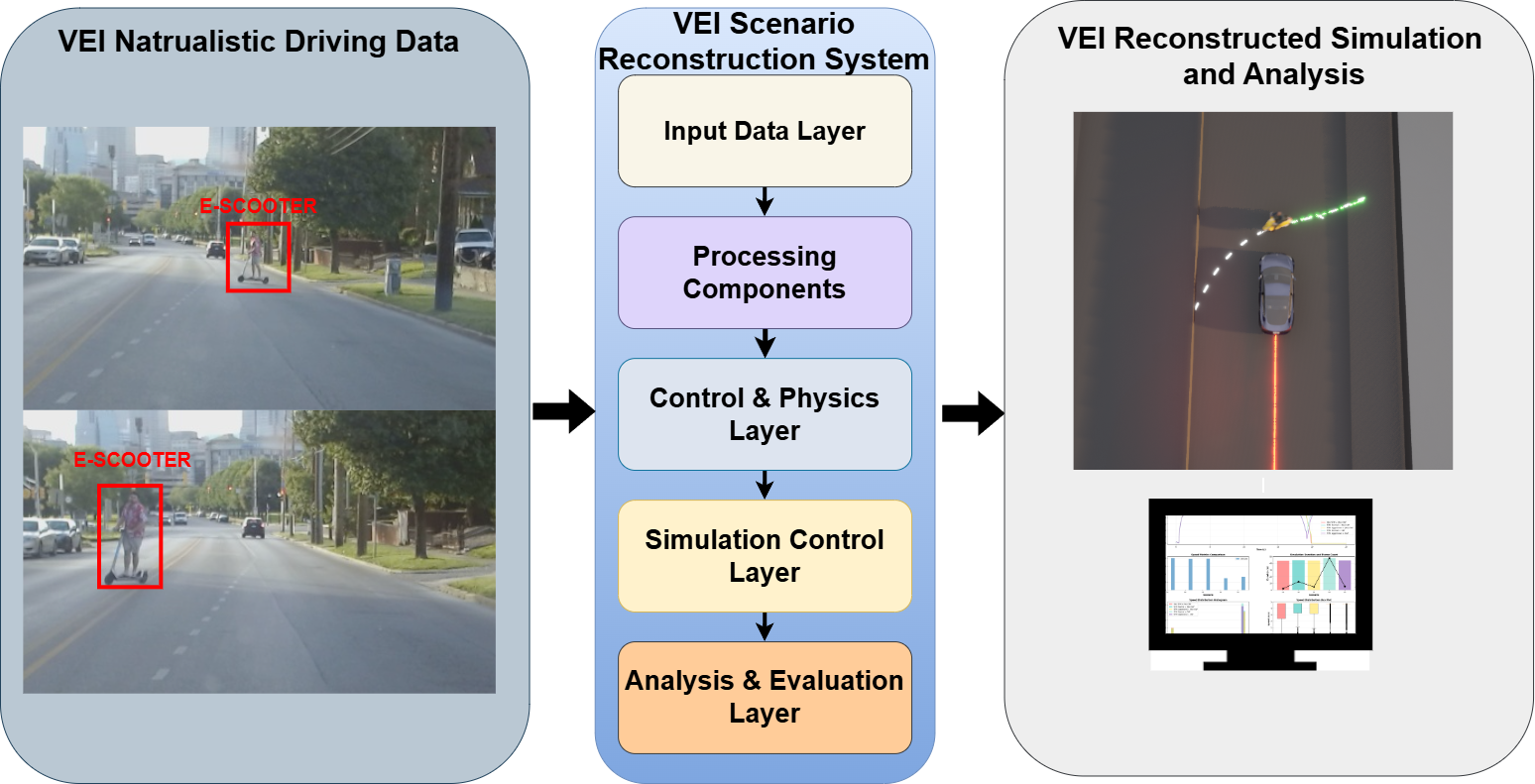}}
\caption{General pipeline for VEI scenario generation from naturalistic driving data.}
\label{fig:vei_intro}
\end{figure}

Building on the collected traffic data that focuses on e-scooter motion behavior by \cite{zhang2025multi}, this paper proposes a pipeline (Fig.~\ref{fig:vei_intro}) that can reproduce real-time vehicle and e-scooter interaction (VEI) in a simulated traffic environment based on the CARLA platform \cite{dosovitskiy2017carla}. Variants of motion behavior for the e-scooter rider are created in the simulation by applying a social force model. The extended synthetic scenarios can be used to test the safety functions, such as emergency braking. An extensive traffic scenario for the VEI can be developed using this pipeline for future AD development.

The key contributions of this paper include:
\begin{enumerate}
    \item Reconstruct the vehicle and e-scooter interaction scenario based on real-world naturalistic driving data.
    \item Provide a configurable and expandable VEI interaction simulation environment.
    \item Demonstrate the vehicular safety feature testing by showcasing some critical scenarios 
\end{enumerate}

The rest of the paper is organized as follows: Section~\ref{Sec:related_works} introduces some recently published dataset containing e-scooter rider and the traffic simulators for AD algorithm validation. Section~\ref{Sec:piepline} illustrates the proposed overall procedure for the VEI scenario reproduction and extension. Section~\ref{Sec:case_study} presents a use case of an example VEI scenario and provides the comparison between a vehicle wiht and without collision avoidance control. The conclusion is drawn in Section~\ref{Sec:conclusion}.

\section{Related Works}
\label{Sec:related_works}

\subsection{Vehicle and E-scooter interaction Study}

Recent studies have conducted interaction analysis through naturalistic datasets and simulation reconstruction. Prabu et al. \cite{prabu2022scendd} presented SceNDD, an urban driving dataset collected with an instrumented vehicle. The paper demonstrated how recorded trajectories can be converted into configurable simulation scenarios through a MATLAB-based pipeline. Ranjan et al. \cite{ranjan2024scenddpp} extended this idea in SceNDD++ to include vulnerable road users such as pedestrians and cyclists. However, their framework does not explicitly model VRU behavior through a dedicated interactive actor.

On the simulation side, Lindner et al. \cite{lindner2022coupled} developed a Unity-based coupled simulator that connects an automated vehicle with a rider-controlled bicycle model. The proposed platform supports human-in-the-loop studies of interaction and communication at conflict points, where simplified vehicle and VRU dynamics are used to maintain simulation synchronization. In \cite{he2023simulation}, a configurable VEI framework with kinematic models for both the ego vehicle and the e-scooter was introduced. Yet the framework was built in a 2D environment and did not explicitly include vehicle perception, which limits sensing-aware control and decision-making studies.

Existing studies show the value of VRU-oriented datasets and simulators for safety analysis. Nevertheless, most focus on pedestrians and bicycles, while e-scooters remain less studied despite their distinct motion characteristics. The reconstruction of real VEI events into high-fidelity simulation environments is still limited. To address this gap, the proposed pipeline aims to reconstruct naturalistic VEI from open data and embed it within a configurable simulation environment that incorporates both perception and motion planning capabilities. The proposed approach also features explicit e-scooter motion models and scenario variations tailored for safety-function testing.

\subsection{Microscopic and Urban Traffic Simulator}

Beyond VRU-focused work, recent studies indicate a growing trend in microscopic traffic simulation and data-driven scenario generation to support the testing and safety analysis of AD functions. Johansson et al. \cite{johansson2023scenario} proposed a scenario-based trajectory generation framework that represents cut-in trajectories using NURBS curves and learns a normalizing flow model over their parameters. The learned model can generate realistic new trajectories and estimate their probability density, which is then used to compute a risk metric for critical scenarios oversampling. In \cite{kolb2024automatically}, Kolb et al. started from recorded urban traffic and derived maneuver-based OpenSCENARIO descriptions. A two-step optimization procedure—continuous parameter tuning was introduced to improve inadequate descriptions automatically.

Other work focuses on scenario cataloging and completeness. Ro{\ss}berg et al. \cite{Roessberg2025assess} utilized a Clustered VQ-VAE (CVQ-VAE) to cluster naturalistic highway scenes into discrete scenario categories, thereby improving codebook utilization and reconstruction quality. Complementing this, a few studies target trajectory reconstruction and imputation using generative models \cite{qian2025diffusion} and microscopic simulation \cite{naing2024fine}. For instance, in \cite{qian2025diffusion}, Qian et al. proposed Diffusion-TGAN, which imputes individual vehicle speeds and reconstructs group longitudinal trajectories from macroscopic speed and spacing measurements. The proposed method produces physically consistent trajectories and significantly reduces speed and trajectory RMSE. Naing et al.,\cite{naing2024fine} formulated trajectory reconstruction using another approach by formulating it as a microscopic traffic simulation–based optimization problem. Traffic simulations were conducted in the SUMO environment, and a trajectory similarity metric was applied. Assignment-based matching between observed and simulated vehicles, along with a data-driven evolutionary optimizer (D3GA++), jointly calibrates model parameters and reconstructs trajectories.

Co-simulation frameworks was developed by Mohammadi et al.,\cite{mohammadi2024sumo2unity} to offer a new scenario-realization pipeline. They introduced a co-simulation tool named SUMO2Unity, which links the microscopic traffic of SUMO with a Unity-based 3D/VR driving simulator via real-time two-way data exchange. Together with SceNDD \cite{prabu2022scendd} and SceNDD++ \cite{ranjan2024scenddpp}, these methods emphasize converting naturalistic logs into editable simulation scenarios.

Overall, these microscopic traffic and scenario-generation studies provide powerful tools for learning, reconstructing, and organizing traffic scenarios. Yet they typically operate on generic vehicle traffic, with limited attention to e-scooter dynamics and behaviors. Few of these studies reconstruct real-world VEI encounters into a modern 3D simulator such as CARLA. To fill the important methodological gap, this paper proposes a method that leverages naturalistic data, an explicit e-scooter motion model, and a CARLA-based simulation environment to generate configurable VEI scenarios that can be used for safety evaluation for the emerging micromobility. 
\section{VEI Scenario Establishment Pipeline}
\label{Sec:piepline}

\begin{figure*}[h]
\centerline{\includegraphics[width=1.0\linewidth]{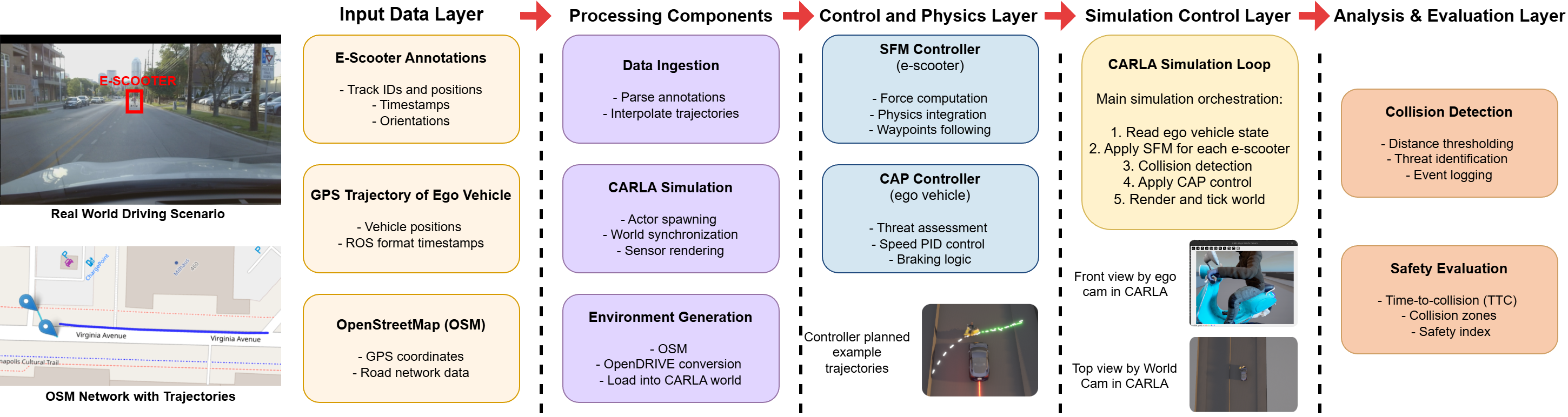}}
\caption{Vehicle and E-Scooter Interaction Scenario Reconstruction System Architecture}
\label{fig:pipeline}
\end{figure*}

\subsection{Data collection}
The VEI scenarios were obtained from real-world data collection activities across the city of Indianapolis, US. A data collection vehicle equipped with front-facing cameras, LIDAR and GPS sensors were utilized to capture driving data of about 10 hours. A joystick based tagging system enabled refining of the raw data to clip and trim segments with e-scooter present on the camera Field of View (FoV). The captured driving data was then processed and manually annotated at a 1 FPS sampling rate (Original data was captured at 10 FPS). Finally, the annotated interactions were formed with lengths ranging from 2 seconds to 30 seconds. We design the VEI pipeline to use these interactions as the input data shown in Fig.~\ref{fig:pipeline}.

\subsection{Scenario Reconstruction}

The pipeline is configured to ingest each input interaction and produce a reconstructed configurable scenario. This scenario would be targeted for a CARLA simulation environment.

The primary steps for the reconstruction process are as follows:
\begin{enumerate}
    \item Firstly, obtain the geographical boundaries of the scenario by analyzing the complete trajectories of the ego vehicle and all e-scooters present in the interaction.
    
    \item Use the boundary coordinates to define the region on Open Street Maps (OSM) and generate the OSM XML metadata.
    
    \item Using the OSM map data, generate the road networks within the CARLA environment by using built-in CARLA  toolkit.
    
    \item Synchronize GPS trajectory timestamps with e-scooter annotation timestamps to ensure frame-by-frame alignment and identify active e-scooters per simulation frame.
    
    \item Transform e-scooter coordinates from ego-relative frame to world coordinates using ego vehicle heading and position, accounting for annotation coordinate convention (Y-axis negation).
    
    \item Compile the trajectories of all agents in the simulations using the annotations, generating sparse keyframes for each tracked e-scooter.
    
    \item Execute cubic spline interpolation of trajectories to generate smooth, continuous motion at simulation frequency, with SLERP-based orientation interpolation.

    \item Finally, configure collision detection metrics to identify the primary threat vehicle for the CAP controller in multi-scooter scenarios.
\end{enumerate}

\subsection{Synthetic Scenario Creation}
For the e-scooter model representation, we use the social force model (SFM) that has been validated as an explicit approahch to mimic the dynamic behaviors of the e-scooter riders\cite{he2023simulation,valero2020adaptation}. The calculated forces will be applied on a point mass model: $\ddot{X_{esc}}=\frac{1}{m_{esc}} f_{total}$,
where ${X_{esc}}\in \mathbb{R}^{2}$ is the e-scooter position state vector and $m_{esc}$ is the e-scooter mass.

\subsubsection{Social Force Model}
\label{sec:sfm}

The total force acting on the e-scooter is decomposed into a destination-seeking component and a vehicle-induced repulsive component:
\begin{equation}
    f_{\text{total}} = f_{\text{des}} + f_{\text{veh}}, 
    \label{eq:ftotal}
\end{equation}
where $f_{\text{des}}$ is a propelling force that drives the e-scooter toward its destination, and $f_{\text{veh}}$ is a repulsive force generated by the ego vehicle.

The destination force $f_{\text{des}} \in \mathbb{R}^{2}$ is defined as
\begin{equation}
    f_{\text{des}} = k_{\text{des}}\bigl(v_{\text{des}} - v_{\text{esc}}\bigr),
    \label{eq:fdes}
\end{equation}
where $k_{\text{des}}$ scales the difference between the desired velocity $v_{\text{des}}$ and the current e-scooter velocity $v_{\text{esc}} \in \mathbb{R}^{2}$. The desired velocity $v_{\text{des}} \in \mathbb{R}^{2}$ is given by $ v_{\text{des}}
    := v_0 \frac{s_{\text{des}} - s_{\text{esc}}} {\left\|s_{\text{des}} - s_{\text{esc}}\right\|^{2} + \sigma_{\text{des}}^{2}},$ where $s_{\text{des}} \in \mathbb{R}^{2}$ is the destination position, $s_{\text{esc}} \in \mathbb{R}^{2}$ is the current e-scooter position, and $\sigma_{\text{des}}$ is a scalar that regularizes the denominator and modulates how the desired speed decays as the e-scooter approaches the destination. The parameter $v_0$ denotes a typical cruising speed for e-scooters.

The repulsive force induced by the ego vehicle, $f_{\text{veh}} \in \mathbb{R}^{2}$, is modeled as
\begin{equation}
    f_{\text{veh}}
    = A_{\text{veh}}\, e^{-b_{\text{veh}}\, d_{v2\text{esc}}}\,
      \vec{n}_{v2\text{esc}}.
    \label{eq:fveh}
\end{equation}

$A_{\text{veh}}$ and $b_{\text{veh}}$ are predefined parameters, $d_{v2\text{esc}}$ is the distance between the ego vehicle’s influential point $s_{\text{influence}}$ and the current e-scooter position $s_{\text{esc}}$, and $\vec{n}_{v2\text{esc}}$ is the unit vector pointing from $s_{\text{influence}}$ to $s_{\text{esc}}$. The exponential term $e^{-b_{\text{veh}} d_{v2\text{esc}}}$ captures a short-range, rapidly decaying repulsion around the ego vehicle.

In the collected VEI data, no collision or near-collision case was found. Therefore, to create a risk-critical traffic sequence, a synthetic simulation was obtained by shifting the timing of the interaction without changing the trajectory. The overlapped trajectories in the spatial-temporal domain indicate a collision. By applying the SFM, an e-scooter can have a dynamic reaction when encountering an approaching motor vehicle. Two types of e-scooter riders were configured: (i) \textit{Aggressive} and (ii) \textit{Normal}. To explicitly set the difference, the \textit{Aggressive} e-scooter only has \(f_{\text{des}}\), indicating a non-cooperative motion behavior. The \textit{Normal} one will keep the \(f_{\text{veh}}\) and react naturally by deviating from the obstacle.

\subsection{Vehicle Collision Avoidance Method}

\begin{figure}[t]
\centerline{\includegraphics[width=0.4\linewidth]{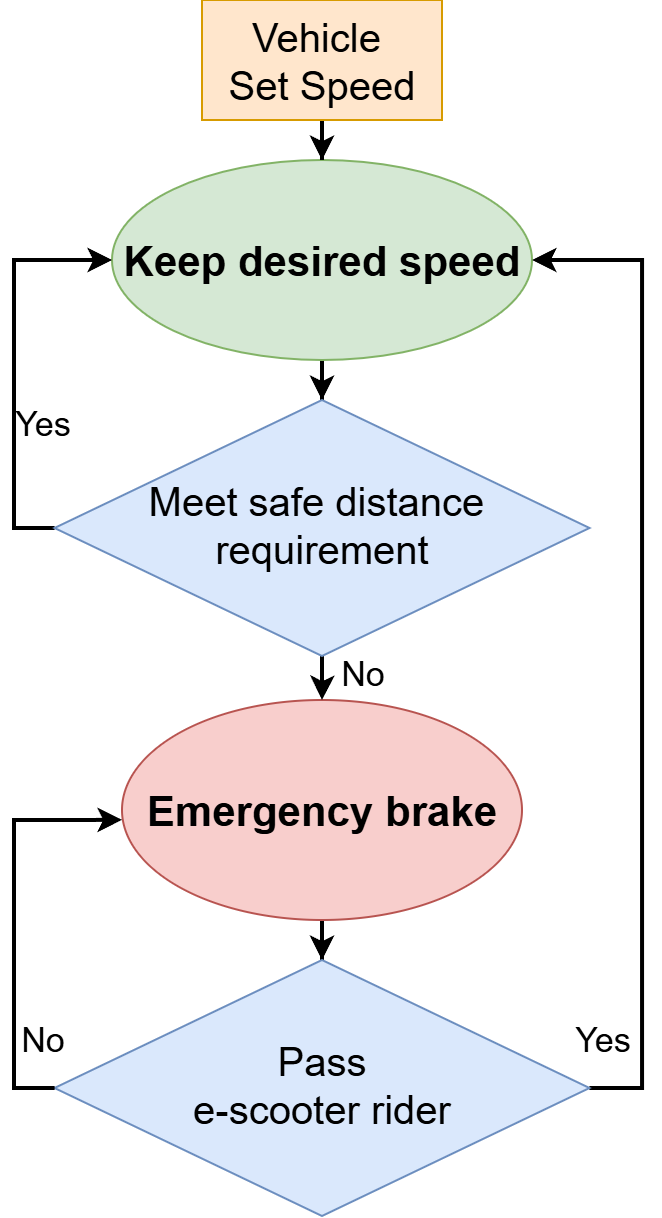}}
\caption{Vehicle Longitudinal Control Flow Chart}
\label{fig:veh_cap}
\end{figure}

A finite state machine, shown in Fig.~\ref{fig:veh_cap}, was developed as an explicit collision avoidance planner. By default, the vehicle under test maintains the desired cruise speed $v_{\text{des}}$ and follows the waypoints processed from the trajectory data. A PID-based longitudinal speed controller was utilized. A distance-based safety layer overrides this behavior when the e-scooter is too close in front of the vehicle.

The speed-dependent safe distance is defined as
\begin{equation}
    d_{\text{safe}}
    = \max\!\left(
        \frac{v_{\text{veh}}^{2}}{2|a_{\min}|},
        T_{\text{safe}}\, v_{\text{veh}}
      \right) + d_{\text{buf}},
    \label{eq:cap_safe_dist}
\end{equation}
where $v_{\text{veh}}$ is the ego speed, $a_{\min}<0$ is the maximum braking deceleration, $T_{\text{safe}}$ is a time–headway
parameter (set to $5\,\mathrm{s}$), and $d_{\text{buf}}$ is a fixed standstill
buffer in front of the e-scooter.

If the e-scooter is in front of the ego vehicle and the
Euclidean gap $d$ is smaller than $d_{\text{safe}}$, an emergency braking command is issued, with the deceleration
saturated at $a_{\min}$. Otherwise, the vehicle continues to follow
$v_{\text{des}}$ under PID control.

\subsection{Evaluation Metric}

We adopt a safety score based on the time-to-collision (TTC) as discussed in \cite{he2024risk} to assess the interaction risk, with
    \(\mathrm{TTC} := \frac{\mathrm{D}_{v2\text{esc}}}{v_{\text{veh}}}\).
In the VEI scenarios considered here, the ego vehicle speed is significantly higher than the e-scooter speed, so we neglect the relative velocity term and approximate TTC using only $v_{\text{veh}}$ in the denominator.

Following the common “two-second rule” \cite{nydmv2secrule}, the TTC time series is partitioned into four zones:
\begin{itemize}
    \item \textit{Safe zone}: $\mathrm{TTC} > 2$, indicating the vehicle maintains a large distance from the e-scooter.
    \item \textit{Attention zone}: $1 < \mathrm{TTC} \leq 2$, where the vehicle–e-scooter gap deserves increased attention.
    \item \textit{Alert zone}: $0 < \mathrm{TTC} \leq 1$, indicating a short time-to-collision and elevated risk.
    \item \textit{Collision zone}: a physical collision occurs between the vehicle and the e-scooter.
\end{itemize}

The safety index is defined as the fraction of time spent in the safe and attention zones:
\vspace{-2mm}
\begin{equation}
    \mathrm{Index}_{\text{safety}}
    := \frac{T_{\text{safe zone}} + T_{\text{attention zone}}}
             {T_{\text{all zones}}},
\end{equation}
where $T_{\text{safe zone}}$ and $T_{\text{attention zone}}$ are the total durations in the corresponding TTC zones, and $T_{\text{all zones}}$ is the total interaction duration.

If a collision occurs at any point during the interaction, we override the metric and set
\(\mathrm{Index}_{\text{safety}}\) as zero to explicitly reflect an undesired outcome.

\section{Case Study}
\label{Sec:case_study}

\subsection{Scenario Illustration}
\begin{figure}[h]
\centerline{\includegraphics[width=0.6\linewidth]{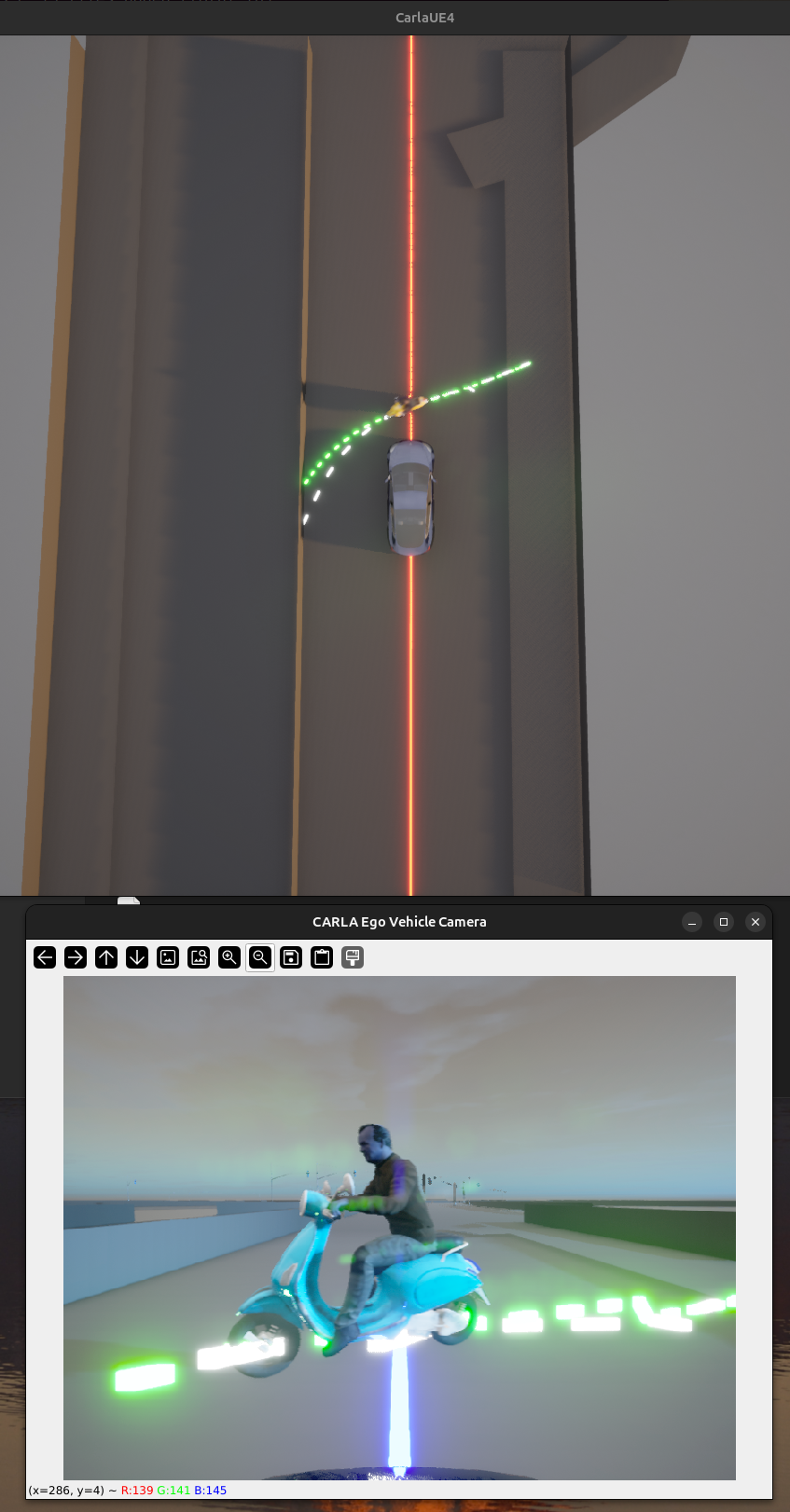}}
\caption{Reconstructed Scenario configured to induce collision (Top: Bird-eye View, Bottom: In-Vehicle Camera view). The trajectory of the ego vehicle is highlighted in solid red while the original waypoints of the e-scooter are marked in dashed white. The dashed green line indicates the actual trajectory of the e-scooter, whose deviation is due to the implementation of the SFM.}
\label{fig:vei_scen}
\end{figure}

\begin{figure}[h]
\centerline{\includegraphics[width=0.7\linewidth]{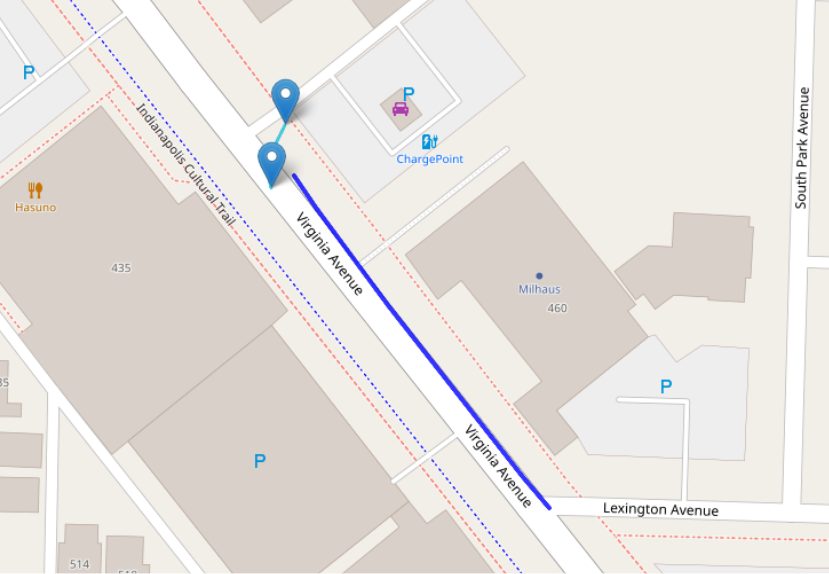}}
\caption{GPS information of the example VEI scenario}
\label{fig:gps_1}
\end{figure}

Fig.~\ref{fig:vei_scen} presents a reconstructed scenario which depicts a VEI scenario on an urban street. The recorded trajectory based on the GPS data is illustrated in Fig.~\ref{fig:gps_1}. 

\subsubsection{Spatial Configuration}
The scenario occurs on a two-lane urban road with clear lane markings (red line indicates ego vehicle heading, green line indicates e-scooter trajectory). The ego vehicle (blue sedan) approaches from behind while the e-scooter crosses laterally, creating a potential collision scenario. 

\subsubsection{Temporal Dynamics}
\begin{itemize}
    \item \textbf{Duration:} 25 seconds
    \item \textbf{Ego vehicle baseline speed:} 5.6 m/s
    \item \textbf{E-scooter nominal speed:} 4.0-4.8 m/s
\end{itemize}

\subsection{Simulation Configuration}

Five distinct simulation variants were designed to isolate the effects of e-scooter behavioral models and vehicle collision avoidance strategies:

\subsubsection{Baseline Configuration}
\begin{itemize}
    \item E-scooter dynamics: Kinematic trajectory replay
    \item Ego control: Constant desired velocity = 5.6 m/s (no avoidance)
    \item Collision detection: Passive monitoring
\end{itemize}

\textbf{Rationale:} Establishes ground truth for comparing active control strategies. Represents scenario replay without behavioral modeling.

\textbf{Expected outcome:} Collision or near-miss event; baseline safety margin measurement.

\subsubsection{Normal E-Scooter Behavior Configuration}
\begin{itemize}
    \item E-scooter dynamics: Social Force Model with normal rider type parameters
    \item Rider type: ``\textit{Normal}'' (moderate avoidance responsiveness)
    \item SFM parameters for Equation~\ref{eq:fdes} and Equation~\ref{eq:fveh}:
    \begin{itemize}
        \item Desired speed gain: $k_{des} = 100.0$ N·s/m
        \item Vehicle repulsion amplitude: $A = 100.0$ N
        \item Repulsion decay rate: $b = 3.5$ m$^{-1}$
        \item Nominal desired velocity: $v_d = 5.0$ m/s
    \end{itemize}
    \item Ego control: Constant velocity (no avoidance)
    \item Collision detection: Passive monitoring
\end{itemize}

\textbf{Rationale:} Isolates e-scooter behavioral response under vehicle threat without vehicle defensive action.

\textbf{Expected outcome:} E-scooter exhibits modest lateral/longitudinal avoidance; ego maintains fixed speed; safety margin improves vs.\ kinematic baseline.

\subsubsection{Aggressive E-Scooter Behavior Configuration}
\begin{itemize}
    \item E-scooter dynamics: Social Force Model with aggressive rider parameters
    \item Rider type: ``\textit{Aggressive}'' (high-gain avoidance)
    \item SFM parameters: Same as normal, but force magnitudes amplified (rider\_type modulation)
    \item Ego control: Constant velocity (no avoidance)
\end{itemize}

\textbf{Rationale:} Tests system robustness to variable e-scooter behavioral aggressiveness.

\textbf{Expected outcome:} E-scooter exhibits less avoidance to vehicle; should exhibit degraded safety margin performance vs.\ normal rider scenario.

\subsubsection{Normal E-Scooter Behavior with CAP Configuration}
\begin{itemize}
    \item E-scooter dynamics: Social Force Model (\textit{Normal} rider)
    \item Ego control: Collision Avoidance Planner (CAP) active
    \item CAP parameters:
    \begin{itemize}
        \item Desired velocity: $v_d = 5.6$ m/s
        \item Safe time-headway: $\tau = 2.0$ s
        \item Deceleration buffer: $d_{buf} = 3.0$ m
        \item PID gains: $K_p=0.05, K_i=0.05, K_d=0.05$
    \end{itemize}
    \item Collision detection: Active threat tracking
    \item Control law:
    \begin{equation}
    a_{\text{cmd}} =
    \begin{cases}
    -\dfrac{v^{2}}{2\bigl(d_{\text{safe}} - d_{\text{buf}}\bigr)},
        & d < d_{\text{safe}},\ \text{VRU ahead}, \\[4pt]
    \mathrm{PID}(v_{d}, v_{\text{current}}),
        & \text{otherwise}.
    \end{cases}
    \end{equation}
\end{itemize}

\textbf{Rationale:} Represents realistic scenario with both e-scooter physics and vehicle defensive control.

\textbf{Expected outcome:} Ego vehicle decelerates in response to detected threat; combined with e-scooter avoidance, achieves maximum safety margin.

\subsubsection{Aggressive E-Scooter Behavior with CAP Configuration}
\begin{itemize}
    \item E-scooter dynamics: Social Force Model (\textit{Aggressive} rider)
    \item Rider type: ``\textit{Aggressive}'' (high-gain avoidance)
    \item SFM parameters: Amplified force magnitudes (Config 3 parameters)
    \item Ego control: Collision Avoidance Planner (CAP) active
    \item CAP parameters: Identical to Config 4
    \item Collision detection: Active threat tracking with aggressive threat model
\end{itemize}

\textbf{Rationale:} Represents worst-case coordination: both e-scooter and 
vehicle employ maximum defensive strategies. Validates system robustness and 
compares coordinated vs.\ solo-controlled responses.

\textbf{Expected outcome:} Earliest avoidance initiation from both agents; 
maximum safety margin; potential over-braking due to dual avoidance signals.





\begin{figure}[b]
\centerline{\includegraphics[width=0.9\linewidth]{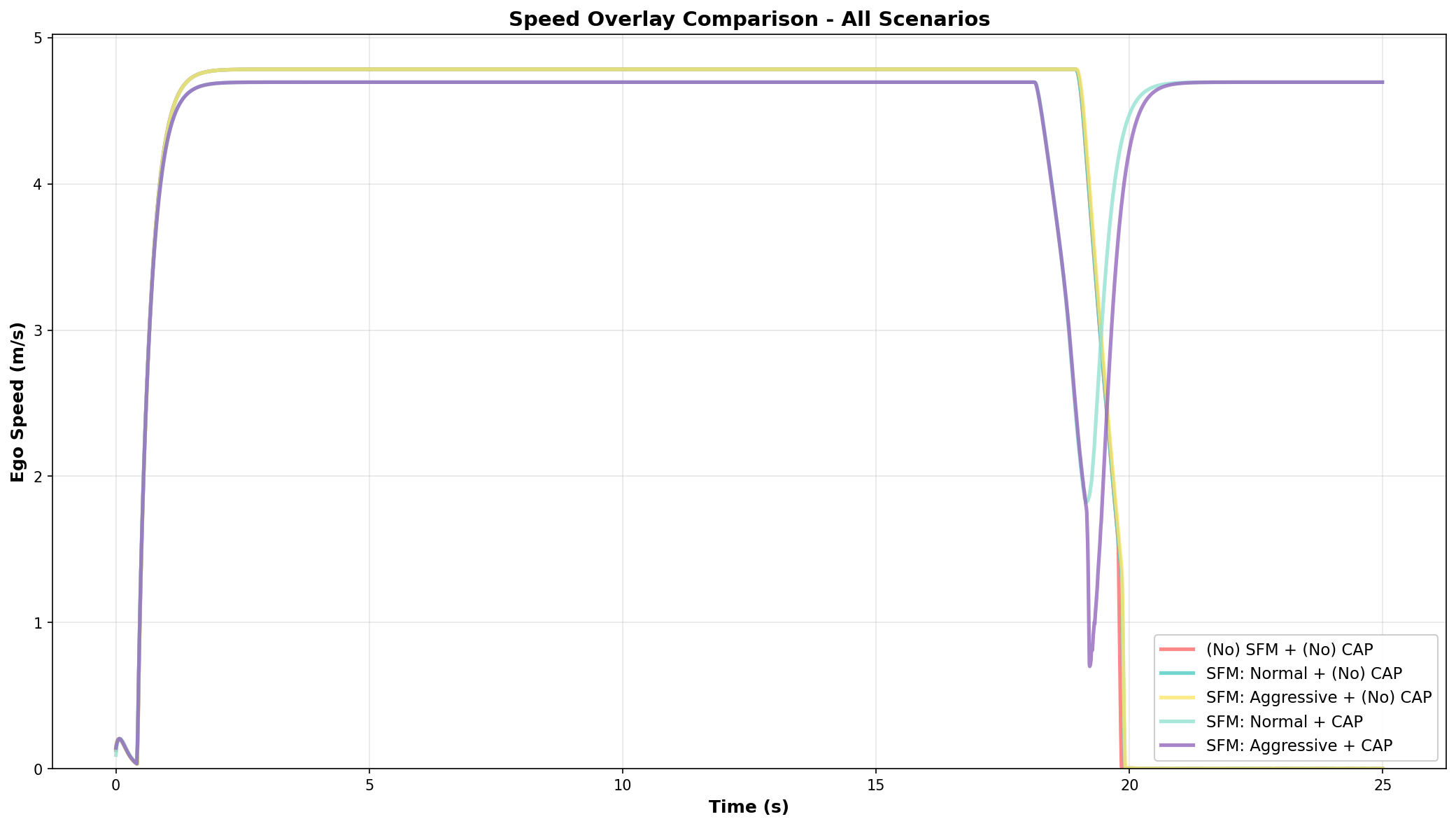}}
\caption{Speed profile for ego vehicle across five synthetic VEI scenarios.}
\label{fig:speed}
\end{figure}

\begin{figure}[h]
\centerline{\includegraphics[width=0.9\linewidth]{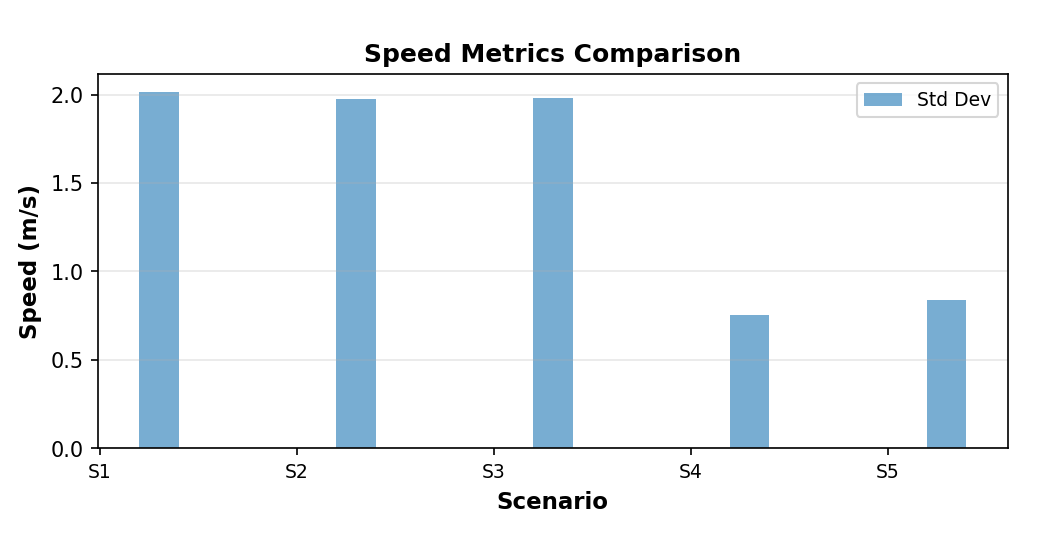}}
\caption{Standard deviation of ego vehicle speed across five scenarios.}
\label{fig:speed_std}
\end{figure}

\subsection{Simulation Results and Discussion}
\label{Sec:discussion}

\begin{figure}[t]
\centerline{\includegraphics[width=0.9\linewidth]{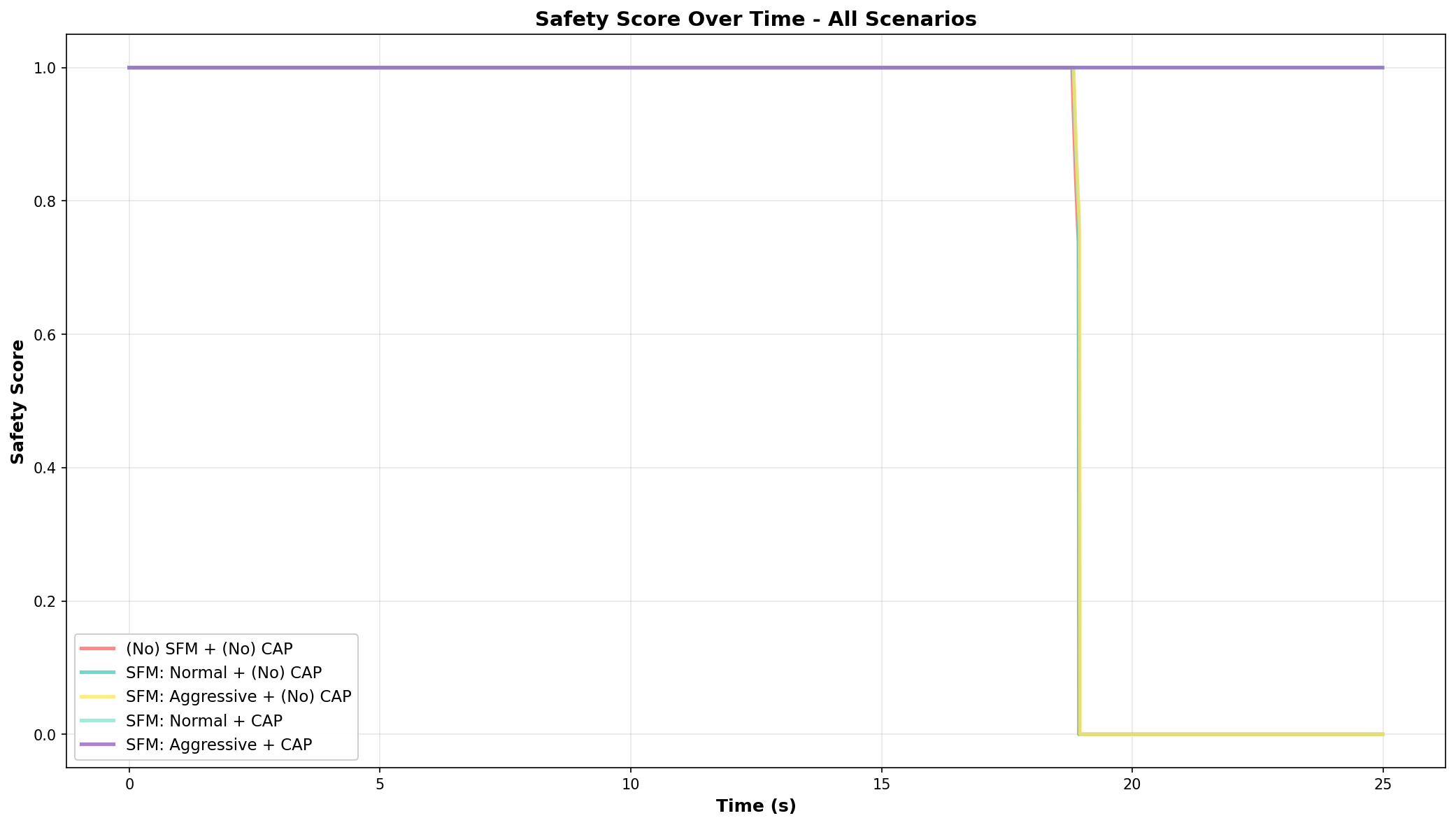}}
\caption{Safety score in the five synthetic VEI scenarios.}
\label{fig:score}
\end{figure}

We evaluate five configurations built from the same reconstructed VEI
baseline in CARLA: (1) baseline configuration without SFM or CAP, (2) \textit{Normal} SFM and no CAP, (3)
\textit{Aggressive} SFM and no CAP, (4) \textit{Normal} SFM with CAP, and (5) \textit{Aggressive} SFM
with CAP. In the first three cases, the ego vehicle tracks the desired
speed and collides with the e-scooter, as reflected by the abrupt drop in
speed to zero in Fig.~\ref{fig:speed} which is further backed by the difference in standard deviation in Fig. \ref{fig:speed_std} . Finally  .The \(\mathrm{Index}_{\text{safety}}\) shown in Fig.~\ref{fig:score}
falling from $1$ to $0$ for the first three configurations also indicates the collision results. This holds when
the \textit{Normal} e-scooter receives a vehicle repulsive force, showing that without a collision avoidance strategy, the vehicle will crash into the evasive micromobility user.

With CAP enabled (configurations~4–5), the ego vehicle starts braking before reaching the e-scooter, and maintains a nonzero gap during the interaction. After the e-scooter is clear and out of the vehicle's feasible path, the ego vehicle accelerates back to the cruise speed. The speed profiles become smoother, and the safety index remains equal to $1$ for the
entire simulation, indicating that the distance-based safety requirement is
never violated for either \textit{Normal} or \textit{Aggressive} SFM behavior.

These results demonstrate the benefits of the proposed pipeline: real-world
VEI encounters from the dataset can be replayed in CARLA to reproduce a
safety-critical traffic event, then systematically extended by (i) shifting interaction timing,
(ii) injecting SFM-based e-scooter dynamics, and (iii) activating CAP to
transform the same reconstructed interaction into a collision-free, safety-compliant
test case for AD functions. 

The proposed methodology can be extended to the safety-critical scenario identification by integrating diverse VEI scenarios. The example scenario only presents the vehicle interaction with a singular e-scooter rider. Multiple e-scooter riders' use cases would exhibit more dynamic interaction behaviors, as different e-scooter riders may result in a much more complex mutual repulsive force. A learning-based method could be implemented to obtain the realistic behavior for multi-e-scooters by training with adequate validated traffic data. Therefore, the established pipeline has huge potential in providing a comprehensive evaluation of a vehicle's safety functionality and investigating the effectiveness of other applications, such as advanced transportation infrastructures.

\section{Conclusions}
\label{Sec:conclusion}

This paper proposes a configurable traffic scenario reconstruction pipeline that focuses on the interaction between vehicles and e-scooter riders. Simulation experiments demonstrate the capability of extending the target scenario to more critical interactions that may result in a potential collision. For future work, the SFM that represents the e-scooter motion behavior can be calibrated using additional public trajectory datasets to model diverse real-world riders. Furthermore, multi-modal sensing based collision avoidance strategies can be incorporated into the pipeline using Vehicle-to-Everything (V2X) communication. The pipeline can be expanded to utilize neural rendering methods to provide a more realistic simulation environment.  Other road users can be integrated into this framework, providing a more comprehensive testing bench for future AD development.

\raggedbottom

\IEEEtriggeratref{17}   

\bibliographystyle{IEEEtran}
\bibliography{root}
	
\end{document}